# On the size of the Fe II emitting region in the AGN Akn 120

Charles A. Kuehn[1], Jack A. Baldwin[1]⋆, Bradley M. Peterson[2] and Kirk T. Korista[3]

**ABSTRACT**

We present a reverberation analysis of the strong, variable optical Fe II emission bands in the spectrum of Akn 120, a low-redshift AGN which is one of the best candidates for such a study. On time scales of several years the Fe II line strengths follow the variations in the continuum strength. However, we are unable to measure a clear reverberation lag time for these Fe II lines on any time scale. This is due to the very broad and flat-topped nature of the Fe II cross correlation functions, as compared to the Hβ response which is much more sharply localized in time. Although there is some suggestion in the light curve of a 300-day response time, our statistical analysis does not pick up such a feature. We conclude that the optical Fe II emission does *not* come from a photoionization-powered region similar in size to the Hβ emitting region, but we cannot say for sure where it does come from. Our results are generally consistent either with emission from a photoionized region several times larger than the Hβ zone, or with emission from gas heated by some other means, perhaps responding only indirectly to the continuum variations.

## 1 INTRODUCTION

The source and excitation mechanism of the strong Fe II emission lines seen in many AGN spectra has been the subject of many papers, but with few definitive conclusions. Wills, Netzer & Wills (1985) showed that the ultraviolet and optical Fe II emission bands carry a substantial fraction of the total energy in AGN emission lines, so that these should not be ignored in any overall model of the Broad Emission Line Region (BELR). The strengths of the optical Fe II emission lines relative to Hβ is an important variable in the Eigenvector 1 found by Boroson & Green (1992), with narrower Hβ lines correlating with stronger optical Fe II emission. This in turn links an understanding of the source of the Fe II emission into an understanding of the nature of Narrow-Lined Seyfert 1 (NLS1) galaxies, and what that may tell us about the basic processes of accretion onto a massive black hole (Collin & Joly 2000; Peterson et al. 2000).

Continuum (Phillips 1978, 1979; Collin et al. 1979, 1980) and line (Netzer & Wills 1983; Sigut & Pradhan 1998; Verner et al. 1999) fluorescence, as well as collisional excitation (Joly 1981, 1987, 1991) have all been proposed as the excitation mechanism for Fe II. So far, no models involving these mechanisms in any combination have really reproduced the full observed Fe II spectrum. Our own recent study of this problem (Baldwin et al. 2004) used a plasma simulation code (Cloudy; Ferland et al. 1998) that includes all of these excitation mechanisms and a 371-level model of the $Fe^+$ ion. We found that a model of a BELR powered by photoionizations from the AGN central engine can satisfactorily reproduce the observed ultraviolet Fe II lines (at wavelengths shortward of Mg II 2800) in terms of both the relative distribution of strengths among the different Fe II lines, and the overall strength of Fe II emission relative to the other strong BELR lines and to the continuum, *provided that* the ionized gas has a strong velocity gradient or large microturbulent velocities. However, this model predicts optical Fe II emission that is an order of magnitude smaller (relative to the UV Fe II lines) than is observed. As an alternative, our study showed that models powered by collisional ionization and excitation could fit the observations equally well, but again could not simultaneously match the UV and optical Fe II emission.

[1] Physics and Astronomy Department, Michigan State University, 3270 Biomedical Physical Sciences Building, East Lansing, MI 48824, USA

[2] Department of Astronomy, The Ohio State University, 140 West 18th Avenue, Columbus, OH 43210, USA

[3] Physics Department, Western Michigan University MS 5252, 1120 Everett Tower, Kalamazoo, MI, USA 49008-5252, USA

⋆ E-mail: baldwin@pa.msu.edu

An observational approach to learning more about the source of the Fe II emission is to try to determine the size of the emitting region by studying the variability of the Fe II line strength. The tool for doing this is the reverberation technique (*e.g.* Blandford & McKee 1982; Peterson 1993). Maoz et al. (1993) successfully measured the reverberation lag time for the Fe II UV emission (covering $\lambda\lambda$2260-3045 Å in the observed frame) in the well-studied Seyfert 1 galaxy NGC 5548, and found it to be about 10 days, similar to that of Ly$\alpha$. This is consistent with the predictions of the photoionization models.

However, attempts to make the same reverberation measurement for the optical Fe II bands to either side of H$\beta$ have not fared so well. These lines clearly do vary in strength in many AGN (Wang, Wei & He 2005, and references therein). In some objects they are observed to vary by a larger amplitude than does H$\beta$, and in other objects by a smaller amplitude. But attempts at determining lag times from reverberation measurements have not been successful. Vestergaard & Peterson (2005; hereafter VP05) very carefully studied the variability of the optical Fe II lines in NGC 5548, but found that the lines were too weak and too heavily blended with lines of other species to obtain anything more than the loose constraint that the variability timescale is less than several weeks. Similarly, Wang et al. (2005) tried to but could not measure a lag time for these lines in the NLS1 galaxy NGC 4051.

In this paper, we attempt to measure the lag time for the optical Fe II lines in the Seyfert 1 galaxy Akn 120. Although this object is not a NLS1 galaxy (the H$\beta$ line has FWHM ~ 5600 km s$^{-1}$), it has quite strong Fe II lines and was the subject of a spectroscopic monitoring campaign over an 8+ year period which yielded a well-determined lag time for H$\beta$ (38.6 days; Peterson et al. 1998a, hereafter P98). Akn 120 would seem to be about as good a prospect as we are going to find for measuring the optical Fe II lag.

## 2 MEASUREMENTS

### 2.1 The data set

We used the same data set that is described by P98. It contains 141 observations of Akn 120 obtained over a 2864 day period with the Ohio State University CCD spectrograph on the 1.8m Perkins Telescope at Lowell Observatory. Our analysis mainly uses the 138 higher-resolution spectra, which have about 10 Å resolution sampled into 2 Å pixels. These cover a wavelength range that varies between observing runs, but extends from about $\lambda$4500-4600 Å at their blue end to about $\lambda$5650-5750 Å at their red end. This corresponds to $\lambda$4355-4451 to $\lambda$5467-5564 Å in the rest frame of Akn 120. In this paper we adopt a redshift of $z$ = 0.03338, measured from a Gaussian fit to the [O III] 5007 emission line in the mean of all 138 spectra. We used the same flux scaling that was used by P98, but that process had trimmed the spectra in wavelength so that they did not fully cover the Fe II bands that we are interested in here. We therefore took the original spectra and scaled them in flux and shifted them slightly in wavelength so that they matched the trimmed spectra.

The observations include nine observing seasons, with gaps in the data during the months when Akn 120 was too close to the Sun to observe. We analyzed the full set of spectra, and then separately the spectra for two well-sampled years which P98 found gave the best measures of the H$\beta$ lag. The two years measured separately are JD48149-48345 (which we designate year 3) and JD49981-50176 (year 8); these are the two years in which the continuum light curve shows the sharpest reversal in slope and thus give the best chance of detecting an optical Fe II reverberation effect on a < 1 yr time-scale similar to that of H$\beta$ (see §4.1). Here and throughout the paper we give the Julian date in a shortened form, after subtracting off 2,400,000 days. The H$\beta$ lag measured for year 8 is the one finally adopted by P98, because it yields the smallest error bars. Our own separate analysis of each of the nine years showed that years 3 and 8 are the ones best suited for the Fe II lag measurements as well.

## 2.2 Light curves

Three major emission features were measured. These were the Fe II bands to the red and blue side of Hβ, and Hβ itself. Figure 1 shows a sample spectrum with the continuum and emission-line wavelength windows marked on it, and the windows are listed in Table 1. These wavelengths and all further wavelengths given in this paper are in the rest frame unless otherwise specified. The spectra do not cover a sufficiently wide wavelength range for us to get a proper estimate of the true continuum level, so we used the same approach as P98 in which we chose pseudo-continuum points in the dips to either side of the emission feature that we were measuring, and then measured the flux above a simple linear interpolation in $F_\lambda$ vs. $\lambda$ between the continuum points. For our Hβ measurement, we used exactly the same windows as P98, and we verified that we got the same results that they did. In the process, we uncovered a minor error in the Julian dates assigned to a few of the year 8 spectra in the P98 analysis, but this does not alter any of the conclusions in that earlier paper.

In their study of the Fe II variability in NGC 5548, VP05 found that the Fe II feature to the blue of Hβ is heavily contaminated by He I λ4471 and He II λ4686 emission lines. Although the Fe II emission is much stronger in Akn 120 than it is in NGC 5548, this contamination is still a major concern. Figure 2 shows a portion of one of the three lower-resolution spectra that we did not include in our main analysis. These spectra included the He I λ5876 emission line. The dashed line in Figure 2 shows the λ5876 line profile moved to the position of λ4471 and scaled on the assumption that $I(\lambda 4471)/I(\lambda 5876) = 0.3$, as was used by VP05 in their analysis. The dashed line also includes a line at the position of He II λ4686, with the profile of the broad feature at this wavelength that shows up in the year 8 rms spectrum described below, but arbitrarily scaled in intensity. The point of Figure 2 is that there is only a narrow wavelength window where the blue-side Fe II emission can be measured without risking strong contamination from helium lines.

Figure 3 shows the measured light curves for these three lines and for the continuum point at λλ5090–5105 Å (the same one used by P98 over the full time span covered by the 138 spectra. The second panel down shows the continuum light curve after applying the correction for possible contamination by broad Fe II emission that is described at the end of §3. For each of the plotted variables, each individual measurement was normalized to the mean value for that variable for all spectra in which it could be measured.

We determined an upper limit on the uncertainty in the $Fe_{blue}$ and $Fe_{red}$ measurements by comparing results from observations separated by four days or less. We find $\sigma(Fe_{blue}) \leq 5.1$ per cent and $\sigma(Fe_{red}) \leq 4.6$ per cent. The flux ratio $Fe_{blue}/Fe_{red}$ is flat over most of the time period, but then gradually changed by about 10 per cent during the last three years. The ratio of continuum level c2 to the other three continuum points shows this same behavior. Since the c1:c3:c4 ratio remains constant, we conclude that the underlying continuum shape has not changed significantly. Rather, there probably is some modest contamination of the c2 continuum point by Hβ and/or He II emission, which in turn contaminates the $Fe_{blue}$ measurement.

## 2.3 RMS and difference spectra

To get a general idea of which features varied in the spectra, we constructed both rms and difference spectra for the full data set and then separately for years 3 and 8. Given the rather perplexing behavior of Fe II that we will discuss below, we show all of these results here. Figure 4 shows the mean and rms spectra for these three samples. We scaled up the rms spectra so that their shapes can more readily be compared to those of the mean spectra. Figure 5 shows the high-state, low-state, and difference spectra. Since we actually are most interested in comparing the Fe II lag time to that of Hβ, we used the Hβ flux to determine high and low states. There is neither evidence nor a theoretical expectation that the optical

Fe II response time should be *shorter* than that of Hβ, so it clearly is more appropriate to use the Hβ flux than the continuum flux. For the full sample, we averaged together all of the year 3 spectra to represent the high state, and all except the first three of the year 8 spectra to represent the low state. While these do not include *all* of the high or low Hβ states in the sample, we believe that they are a fair representation of them, and they have the practical advantage that they come from the portion of the data set on which we have done the most work. For the separate analyses of years 3 and 8, we averaged together the three spectra with highest Hβ fluxes to create the high state spectrum, and those with the three lowest Hβ fluxes for the low state.

The rms and difference spectra for year 8 both show a relatively strong broad feature centered at about rest wavelength λ4671 Å. We interpret this as a variable component of the He II λ4686 line.

We checked carefully for any contamination of the $Fe_{blue}$ or $Fe_{red}$ bumps by any non-varying narrow features. Figure 6 shows the high- and low-state spectra for each year, after a continuum fit has been subtracted. In each panel, the high state has been scaled to match the low state at λ5300 Å. Clearly, except for the He II λ4686 lines, all parts of the Fe II bumps vary by the same amount. Also, the two Fe II bands and Hβ have varied by nearly identical fractional amounts in all three cases.

## 3 LIGHT CURVE ANALYSIS

The light curves for the three emission features were analyzed using standard techniques to cross correlate them with the continuum light curve. The model-independent Monte-Carlo technique (Peterson et al. 1998b; Peterson et al. 2004) was used with interpolation cross correlation functions (ICCFs: Gaskell & Sparke 1986; Gaskell & Peterson 1987) to estimate, for the years 3 and 8 data subsets and for the full data set, the best-fitting lag time and its uncertainties. We calculated lag times τ for both the peak and the centroid of the cross-correlation functions. The results are given in Table 2, along with the Hβ results for these from P98. The $τ_{centroid}$ measurement is generally the more robust of the two, and it is clear that our $τ_{centroid}$ results for Hβ are in good agreement with those of P98. Since we used the same data set as P98, this merely verifies that the slight modifications to the analysis technique over the years have not changed the result. The rather large difference in $τ_{peak}$(Hβ) for year 8 reflects the error in assigning Julian dates that was mentioned above.

Table 2 also shows that we could not obtain a meaningful measurement of the lag for either of the Fe II features in either of the individual years. The Monte-Carlo technique produces cross-correlation centroid distributions which show the probability of each lag time being the correct one. These functions are sharply peaked for Hβ, but are broad with multiple peaks for $Fe_{blue}$ and $Fe_{red}$. The large error bars on the Fe II lags reflects this structure in their cross-correlation centroid distributions. Such behavior is caused by aliasing. Although this often is due to poor time spacing of the observations, in this case the time sequence within the individual years is well-sampled. The problem instead arises from the shape of the cross-correlation functions (CCFs). The CCFs for Fe II and Hβ are shown in Figure 7, for years 3 and 8 and also for the full data set. The solid lines are the results using the ICCF method, while the individual points were calculated using the discrete correlation function (DCF) method of Edelson & Krolik (1988). The Hβ CCF reaches a clear peak for all three data sets, while the $Fe_{blue}$ and $Fe_{red}$ CCFs are extremely broad and flat-topped. Simply put, the optical Fe II lines do not seem to reverberate as cleanly as do other lines.

Visual inspection of the light curves shown in Figure 3 suggests that the Fe II lines vary with a time lag of about 300 days. The Fe II lag times listed in Table 2, measured from the interpolation CCFs computed over all years, at first seem to confirm this. However, repeating the Monte Carlo probability analysis using the DCF technique shows no sign of these lags. This indicates that the ICCF results probably are spurious, and are due to the low amplitude of the Fe II CCFs in combination with the gaps in the sampling and noise in the measurements.

There is a potential for problems due to contamination of the measured continuum points by the Fe II "pseudo-continuum" (caused by the wings of the individual Fe II lines blending together). This is often dealt with by constructing template profiles of the Fe II bands and then fitting them to the data. However, as is shown in the following paragraphs, the resulting continuum flux depends critically on the actual shape of the profiles of the individual Fe II lines that make up the broad bands, which we simply do not know.

To investigate the potential importance of contamination of the continuum points, we found a "worst case" by taking a wide variety of plausible Fe II profiles and convolving them with an Fe II template (from Véron-Cety, Joly & Véron 2004, for the broader lines in I Zw 1). We tried ten different Hβ profiles as the convolution kernel used to create the smoothed template. These included a Gaussian with the same FWHM (5600 km s$^{-1}$) as the Hβ line in the mean spectrum for all years. Figure 8 shows three examples of the results. The Hβ profile in the upper panel is meant to represent the full Hβ profile in the mean spectrum for all years. It was created by first subtracting a narrow Hβ spike with an intensity 1/12 that of [O III] λ5007, which is the generally accepted contribution to Hβ from the NLR in this object (Korista 1992). We fitted a smooth quadratic continuum through points c1, c2, c3, c4, and then sketched in a smooth Hβ wing under the red wing of He II λ4686. Finally, to estimate the Hβ contribution in the region under the [O III] λλ4959, 5007 lines, we reflected the blue wing of Hβ around zero velocity and used it as the red wing of Hβ at wavelengths beyond λ4940 Å. The Hβ profile in the middle panel of Figure 8 is from the rms spectrum for all years (see Fig. 4), and represents the variable Hβ component. The "constant peak" profile in the lower panel was created by taking just the narrow central peak left behind when the rms Hβ profile from the center panel was scaled to match the one in the upper panel and then subtracted. It is intended to represent some very slowly varying low velocity component in the BELR, for purposes of experimenting with the effects of using different convolution kernels.

As a worst-case pseudo-continuum contribution, we took the strength of the smoothed Fe II template at the wavelength of the c3 continuum point (indicated in Fig. 8 by the dotted vertical line at 5100 Å), for the particular case shown in the upper panel of Figure 8. That profile produced the largest pseudo-continuum contribution of any of the ones we tried. The pseudo-continuum contribution from the smoothed template is directly proportional to the total flux in that template, and therefore directly proportional to the fraction of that total flux which is included in our Fe$_{red}$ measurement. Therefore, we can make a correction to the c3 continuum level measured in each individual spectrum by multiplying the appropriate proportionality constant by the Fe$_{red}$ flux measured in that spectrum. Subtracting that correction from the measured c3 continuum level is equivalent to fitting the smoothed Fe II template to each individual spectrum. Since we don't really know which smoothed template to use, there is little point in carrying out a full fitting procedure. The corrected c3 continuum light curve is shown in the second panel down in Figure 3. We reran the programs for measuring lag times using these corrected continuum levels, and found that it made only small differences in any of our results, amounting to an increase in the lag times of 10 per cent or less in most cases. These are not significant in comparison to the error bars on the lag measurements. In view of the fact that we have no real idea of which smoothed Fe II template to use, we use the results from the uncorrected continuum points.

We note that the situation for Akn 120 is different than VP05 encountered in their analysis of NGC 5548. For NGC 5548, the Fe II lines are extremely weak but the wavelength coverage was broad enough so that the continuum shape could be measured to reasonable accuracy, so fitting template Fe II profiles was helpful. For Akn 120, we do not have enough wavelength coverage to properly measure the underlying continuum level as a prelude to fitting a template profile, but the Fe II lines are so strong that there is no real need for fitting a template.

## 4 DISCUSSION

### 4.1 Optical Fe II variability relative to the continuum

We showed in Section 3 that the optical Fe II emission is correlated with the continuum over long time scales, but an analysis of these light curves shows that Fe II definitely does not vary in the same way as Hβ. Visual inspection of Figure 3 does suggest a possible Fe II lag time of about 1 year, but our statistical analysis of the light curve is unable to detect it.

To identify a lag, the cross correlation function requires a change in sign of the continuum and line variations as a function of time after passing through a maximum or minimum in the light curves, a "hook" in the light curves. Such a "hook" is apparent during the first ~ 1000 days of the campaign in which the optical Fe II light curve appears to lag behind the optical continuum by ~ 1 year. Most of the remaining light curve is a long downward trend with very little "hook" in it, and so adds very little power to the CCF except at zero time-lag. In fact within any individual observing season, the optical Fe II light curves are largely featureless. This is true even in year 8, during which the continuum underwent a significant reversal in flux (see Figure 3). Unlike the case for broad Hβ, the measured optical Fe II emission line light curves do not show correlation with the continuum except possibly on the longer time scales for which the continuum variations have their largest power. Although this could be due in part to difficulties in measuring the broad, low-contrast Fe II features, the error bars in Figure 7 indicate that we should have been able to detect an Fe II response as weak as about half that of Hβ. We believe that instead the problem is caused by the intrinsic nature of the optical Fe II line-emitting region (e.g., modest line responsivity $dlogF_{line}/dlogF_{cont}$ with respect to the incident continuum variability, perhaps coupled with large geometric dilution).

**4.2 Fe II emission from an extended BELR?**

One possible way to produce the very broad CCFs seen for Fe II is for the Fe II emitting region to be directly photoionized, but to be much more extended than the region that emits Hβ. One check of whether or not this is plausible is to try to estimate the widths of the individual Fe II lines, and compare them to the width of Hβ. It has been shown in the handful of AGN for which lags have been measured for multiple lines (Peterson & Wandel 1999, 2000; Onken & Peterson 2002; Kollatschny 2003; Metzroth, Onken & Peterson 2006) that the line widths $v$ of the strong emission lines are correlated with the sizes $r$ determined from the lag times of these lines in the way expected for virialized gas motion, $v^2r = constant$. If Fe II does in fact have a lag time of roughly 300 days, it comes from a region approximately six times larger than Hβ. We would therefore expect the Fe II lines to be about $6^{1/2}$ times narrower than Hβ, or $FWHM_{Fe\,II}$ ~ 2300 km s$^{-1}$. Figure 8 compares the observed Fe II spectrum to a variety of broadened Fe II templates. The simulation using the full Hβ profile (upper panel) appears smoother than the observed spectra, but best reproduces the overall shapes of the Fe II bumps, while the two simulations using narrower line profiles do a better job of producing sharp narrow sub-peaks like those on top of the observed Fe II bumps. It is noticeable that the lines in the original Fe II template (the series of sharp spikes in the upper panel of Figure 8) do not actually line up with the small peaks on top of the observed Fe II bands, so apparently the template is not a very accurate description of the unbroadened Fe II spectrum in Akn 120 in any case. We conclude that observed structure in the Fe II bumps is consistent with their being emitted from anywhere between the same radius as the Hβ emission to several times farther out.

A concern about a very large photoionized Fe II emission region is that there should be a maximum extent to the BELR, defined by a dust sublimation radius $R_{dust}$ (Netzer & Laor, 1993). While dust is not expected to survive against sublimation within the highly ionized parts of the BELR, it should be present in large quantities at radii starting somewhere beyond the region where most of the Hβ is formed. This would shield any extended BELR gas from ionizing radiation. According to Elitzur & Shlosman (2006), for a typical AGN ionizing continuum shape $R_{dust} = 0.4\,L_{45}^{1/2}$ pc, where the grain temperature drops to 1500 K. Here $L_{45}$ is the bolometric luminosity in units $10^{45}$ erg s$^{-1}$. We use the rule of thumb that $L_{bol} = 10\,\lambda F_\lambda(5100\text{Å})$, which gives $L_{45} = 0.8$ and $R_{dust} = 0.36$ pc = 430 light days for the observed mean $F_\lambda(5100\text{Å})$ for Akn 120. Given the uncertainties, $R_{dust}$ is the same as the radius indicated by the possible 300-day Fe II lag time, so it is plausible that the optical Fe II emission is produced at or just inside the dust sublimation radius.

The main argument against having an Fe II-emitting region that is just an extension of the BELR is that the observed Fe II emission is much stronger than is predicted by photoionized models. Using the smoothed Fe II template fits shown in Fig. 8 to estimate the true continuum level in the mean spectrum for all data, we find $I$(Fe II$_{optical}$) $/I$(H$\beta$) ~ 0.7–1.4 and an equivalent width $W_\lambda$(Fe II$_{optical}$) ~ 100–200 Å, depending on which velocity profile is assumed for the Fe II lines. The Fe II$_{optical}$ /H$\beta$ intensity ratio predicted by the Baldwin et al. (2004) models that fit the observed UV bump in other AGN is 5–10 times lower than the value measured for Akn 120. The Baldwin et al. model grids produce an optical Fe II equivalent width as high as the observed value in just one corner of the (ionizing flux, gas density, column density) parameter space studied in that paper, corresponding to fairly low density gas ($n_H \leq 10^9$ cm$^{-3}$) subjected to low ionizing flux. When the additional regions of parameter space needed to produce the other observed emission lines (including H$\beta$) are also counted in, the integrated Fe II equivalent width from the models is far lower than the observed value for Akn 120. This all shows that the optical Fe II emission from Akn 120 is roughly an order of magnitude stronger than that predicted by photoionization models, as is the case with other QSOs with strong Fe II lines in the optical passband. This could mean either that the models are incomplete, or that the optical Fe II emission is not directly powered by photoionizations.

### 4.3 Optical vs. UV Fe II emission

It is possible that the optical and UV Fe II lines come from quite different regions. In NGC 5548, the one case where it has been measured (Maoz et al. 1993), the Fe II UV bump shows reverberation response with a time lag similar to that of C IV and Ly$\alpha$. However, a very thorough attempt (VP05) to measure reverberation behavior of the optical Fe II lines in this same object ran into the same sort problems with ill-defined CCFs that we encounter here with Akn 120. The optical Fe II lines in NGC 5548 are quite weak, so no firm conclusion could be reached about their reverberation behavior, but their response does seem to different than that of the Fe II UV bump, suggesting that they might come from a very different region of the gas.

We do not have simultaneous measurements of the UV and optical parts of the Fe II spectrum for Akn 120, so their flux ratio is not known. However, taking the NGC 5548 results together with the present result for Akn 120 suggests that most of the energy in the UV Fe II lines might come from the comparatively higher ionization region that produces most of the C IV and Ly$\alpha$ emission, while the optical Fe II arises, as has been widely speculated, come from a different, lower-ionization region.

### 4.4 Collisionally excited models

The above discussion implies that there may exist a region that produces strong Fe II optical emission bands, but little emission from the Fe II UV bump or from lines of other ions. An alternative to photoionization is that the optical Fe II emission is powered by collisional excitation. Collin-Souffrin and her co-workers (e.g. Joly 1987; Dumont, Collin-Souffrin & Nazarova 1998; see also Kwan et al. 1995) have proposed that this occurs in gas in an accretion disk rather than in the photoionization-dominated part of the BELR. In such a situation, the Fe II strength could still correlate with the continuum strength in some indirect way, since the continuum emission is also produced by processes connected with the accretion disk.

The best calculations to date of this sort of model (Baldwin et al. 2004) do not fit the full observed Fe II spectrum. Gas with temperature $5000 \leq T_e \leq 20,000$ K, density $n_H$ ~ $10^{12}$–$10^{16}$ cm$^{-3}$, and column density $N_H$ ~ $10^{25}$ cm$^{-2}$ does emit primarily Fe II lines, and can reproduce the observed shape and strength of the $\lambda\lambda$2240–2650 Å UV bump. However, the computed optical emission from these models does not match the observed spectrum. Low density gas produces most of its Fe II emission in the optical bands, but the observed multiplets 37 and 38 (which makes up the Fe$_{blue}$ feature discussed here) are missing in the model spectrum. Higher density models with $n_H = 10^{14}$ or $10^{16}$ cm$^{-3}$ produce optical Fe II emission with relative strengths much more like what is observed, but also produce UV emission lines vastly stronger than is observed, and which occur as a pair strong narrow spikes in the UV spectrum rather than the observed broad bump.

The models described in the preceding paragraph assumed that the turbulent velocity $v_{turb} = 0$. In the light of our Akn 120 results, we ran additional models with $v_{turb} = 100$ km s$^{-1}$, to see if we could get a better match to the observed spectra of the optical Fe II lines within the constraints of not overproducing the Fe II UV bump or the lines of other ions. We used CLOUDY (Ferland et al. 1998) to run a sequence of collisionally excited models with column density $N_H = 10^{25}$ cm$^{-2}$, $T_e = 6340$ K, $v_{turb} = 0$ and 100 km s$^{-1}$, and hydrogen density $n_H = 10^{10}$, $10^{12}$, $10^{14}$ and $10^{15}$ cm$^{-3}$. The models with $v_{turb} = 0$ km s$^{-1}$ served to verify that we got the same results as in the Baldwin et al (2004) paper. Figure 9 shows the results. The first two columns show the predicted Fe II emission from the $v_{turb} = 0$ and 100 km s$^{-1}$ models, respectively, with the computed spectra binned into 580 km s$^{-1}$ intervals. The presence of microturbulence clearly strengthens the optical Fe II lines relative the UV bump. The right-hand column in Figure 9 shows the $v_{turb} = 100$ km s$^{-1}$ models convolved with a Gaussian profile with $FWHM = 5600$ km s$^{-1}$, to match the H$\beta$ line width in Akn 120.

However, Figure 10 shows that while the broadened $10^{10}$ cm$^{-3}$ model matches the observed Fe II spectrum to the red of H$\beta$, it dismally fails to reproduce the observed spectrum to the blue of H$\beta$. As had previously been found for the $v_{turb} = 0$ models in this same density range, the bump visible near $\lambda$4300 Å in the upper right-hand panel of Figure 10 is at a shorter wavelength than the observed bump. The higher density models produce the correct shape for the Fe II emission, but predict strong Ca II H and K lines which are not observed in the spectra of the real AGN. They also predict Mg II $\lambda$2800 to be one of the strongest lines from this region. Mg II is in fact strong in AGN spectra, but at least in NGC 4151 (one of the few cases where it has been measured; Metzroth, Onken & Peterson 2006), it reverberates on a well-defined timescale, unlike the optical Fe II lines in Akn 120 and NGC 5548. In addition, these higher density models would produce the Fe II UV bump from the same gas that produces the Fe II optical lines, contrary to the possible difference in UV and optical Fe II lag times discussed above. Thus, none of the collisionally excited models which we ran for this study actually fit the data. Perhaps there are other parameters for such models that would produce better fits, but finding them is beyond the scope of the current paper.

## 5 RESULTS

The optical Fe II lines do not reverberate on the same 40 – 50 day timescale as the H$\beta$ line. Although the general appearance of the light curves suggest that Fe II might reverberate with a much longer lag time, we find the low amplitude and long duration of the apparent Fe II response to be disturbing. The cross correlation functions are very broad and flat-topped, as opposed to the H$\beta$ response which is much more sharply localized in time. As a result, our analysis of the light curve is unable to pick up any clear-cut reverberation signal.

In spite of the lack of any real detection of a lag time, our results still are consistent with the possibility that the optical Fe II lines come from an outer part of the photoionized BELR, reaching out to the edge of the BELR, which we assume is set by the dust sublimation radius at $R_{dust} \sim 430$ light days. Tests in which we broadened an Fe II template spectrum by various amounts indicate widths of the individual Fe II lines may be anywhere from the same as for the full observed H$\beta$ to a fraction of that width (corresponding to a considerably larger radius for virialized motions).

Current models of photoionized BELR gas do not predict this sort of extended region producing the optical Fe II emission lines. At the same time, these models are able to explain the Fe II UV bump as coming come from a photoionized gas that has strong velocity gradients due either turbulence or some sort of flow, and which lies in the main part of the BELR. This suggests that the optical Fe II lines do not come from the same region that produces the Fe II UV bump.

Our attempt to model the optical Fe II lines as being due to a collisionally ionized, turbulent gas component (which might, for example, be part of an accretion disk) was not successful. These models produced strong emission lines of other ions which are not present in observed AGN spectra, and they do not fit the observed shapes of the Fe II bumps very well.

**ACKNOWLEDGEMENTS**

We gratefully acknowledge support from NSF through grant AST-0305833 and from NASA through HST grant GO09885.01-A to MSU and grant AST-0604066 to OSU.

**FIGURE CAPTIONS**

**Figure 1.** Sample spectrum, showing the four wavelength ranges used to define the pseudo-continuum levels c1, c2, c3 and c4, and the three ranges ($Fe_{blue}$, H$\beta$ and $Fe_{red}$) in which the emission line fluxes above the interpolated continuum were summed.

**Figure 2.** Low resolution spectrum from JD47777, showing a synthetic blend of He I $\lambda$4471 and He II $\lambda$4686 lines and the position of the $Fe_{blue}$ wavelength window. The profile and strength of the $\lambda$4471 line was scaled from the measured He I $\lambda$5876 line, assuming $I(\lambda 4471)/I(\lambda 5876) = 0.3$. The He II profile is that of the broad feature which shows up in the Year 8 rms spectrum (Fig. 4), with an arbitrary scaling in flux.

**Figure 3.** Light curves for the 5100Å (c3) continuum point and for the H$\beta$, $Fe_{blue}$ and $Fe_{red}$ emission line fluxes, normalized by their mean values. The solid lines in the lower four panels are the measured continuum light curve, shown for reference. The "Corrected Continuum" points in the second panel down show the (small) effect of correcting the c3 continuum measurements for contamination by the pseudo-continuum due to the Fe II emission bands, using the method explained at the end of §3.

**Figure 4.** Mean and rms spectra for the full data set, and separately for years 3 and 8. The rms spectra have been scaled in flux by a factor of 4, to make them easier to see.

**Figure 5.** High, low and difference spectra for the full data set, and then separately for years 3 and 8.

**Figure 6.** High and low-state spectra after the continuum has been subtracted, and then the remainders have been scaled to match at $\lambda$5300Å. This shows that the shapes of the two Fe II emission bumps have not changed (except in the He II $\lambda$4686 region for Year 8).

**Figure 7.** Cross-correlation functions for H$\beta$, $Fe_{blue}$, and $Fe_{red}$ for all years combined, and separately for years 3 and 8. In each frame the interpolation cross-correlation function is shown as a solid line, and the discrete cross-correlation function as individual points with error bars.

**Figure 8.** The observed mean spectrum for all years (upper solid line in upper panel), compared to examples of smoothed template spectra (remaining solid lines in the three panels). The smoothed template spectra were created by convolving various possible H$\beta$ profiles (dashed lines), created as described in §3, with the Fe II template profile from Véron-Cety et al (2004) which is shown as the series narrow spikes in the upper panel. The "Fe II smoothed" spectrum from the upper panel was used as our estimate of the maximum Fe II contribution to the measured c3 continuum point. The "Corrected Continuum" light curve shown in Figure 3 was computed by scaling this Fe II "pseudo-continuum" contribution in direct proportion to the measured $Fe_{red}$ flux, and then subtracting it from the measured continuum value.

**Figure 9.** Collisionally excited models with $T_e = 6340$ K. The left hand column shows models with $v_{turb} = 0$, for a sequence of gas densities $n_H = 10^{10}, 10^{12}, 10^{14}, 10^{15}$ cm$^{-3}$ (top to bottom). The middle column shows the same models but with $v_{turb} = 100$ km s$^{-1}$. The right-hand column shows the models from the middle column after convolution with a 5600 km s$^{-1}$ FWHM Gaussian profile to match the measured H$\beta$ line width.

**Figure 10.** Comparison of the observed spectrum to that of a collisionally excited model with $T_e$ = 6340 K, $n_H$ = 10$^{10}$ cm$^{-3}$, $v_{turb}$ = 100 km s$^{-1}$, which has been convolved with a FWHM = 5600 km s$^{-1}$ Gaussian profile. This illustrates how the model provides a satisfactory fit to the lines to the red of H$\beta$, but not to those to the blue.

| Table 1 | | |
|---|---|---|
| Continuum and Emission Line Measurement Windows | | |
| Continuum | Rest Wavelength Range (Å) | Observed Wavelength Range (Å) |
| c1 | 4451–4464 | 4600–4613 |
| c2 | 4737–4756 | 4895–4915 |
| c3 | 5090–5105 | 5260–5275 |
| c4 | 5438–5458 | 5620–5640 |
| Emission Line | | |
| $Fe_{blue}$ | 4548–4587 | 4700–4740 |
| H$\beta$ | 4756–4940 | 4915–5105 |
| $Fe_{red}$ | 5129–5322 | 5300–5500 |

| Table 2 | | | | | | |
|---|---|---|---|---|---|---|
| Cross-Correlation Results | | | | | | |
| Line | $\tau_{peak}$ (days) | | | $\tau_{centroid}$ (days) | | |
| | All years | Year 3 | Year 8 | All years | Year 3 | Year 8 |
| $Fe_{blue}$ | 303 ± 135 | -13.6 ± 63.0 | 35.0 ± 42.7 | 313 ± 73 | -3.7 ± 55.7 | 25.3 ± 36.7 |
| $Fe_{red}$ | 314 ± 96 | 25.7 ± 52.8 | 54.7 ± 53.2 | 341 ± 41 | 35.3 ± 38.8 | 29.9 ± 52.6 |
| H$\beta$ | 44.1 ± 14.5 | 49.3 ± 25.9 | 40.0 ± 11.5 | 57.3 ± 23.0 | 49.7 ± 17.3 | 38.6 ± 7.4 |
| H$\beta$ (from P98a) | $52^{+2}_{-14}$ | $54^{+11}_{-16}$ | $29^{+24}_{-1}$ | $60.2^{+31.1}_{-13.2}$ | $49.5^{+12.4}_{-14.6}$ | $38.6^{+5.3}_{-6.5}$ |

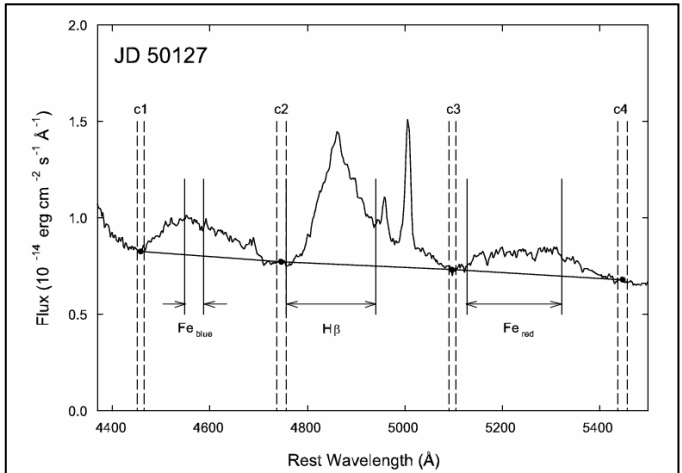

**Figure 1.** Sample spectrum, showing the four wavelength ranges used to define the pseudo-continuum levels c1, c2, c3 and c4, and the three ranges (Fe$_{blue}$, H$\beta$ and Fe$_{red}$) in which the emission line fluxes above the interpolated continuum were summed.

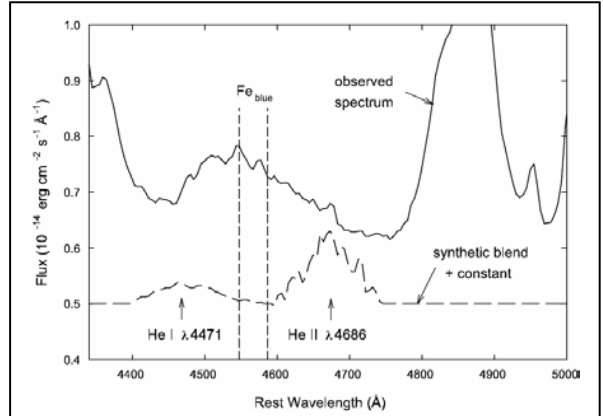

**Figure 2.** Low resolution spectrum from JD47777, showing a synthetic blend of He I $\lambda$4471 and He II $\lambda$4686 lines and the position of the Fe$_{blue}$ wavelength window. The profile and strength of the $\lambda$4471 line was scaled from the measured He I $\lambda$5876 line, assuming $I(\lambda 4471)/I(\lambda 5876) = 0.3$. The He II profile is that of the broad feature which shows up in the Year 8 rms spectrum (Fig. 4), with an arbitrary scaling in flux.

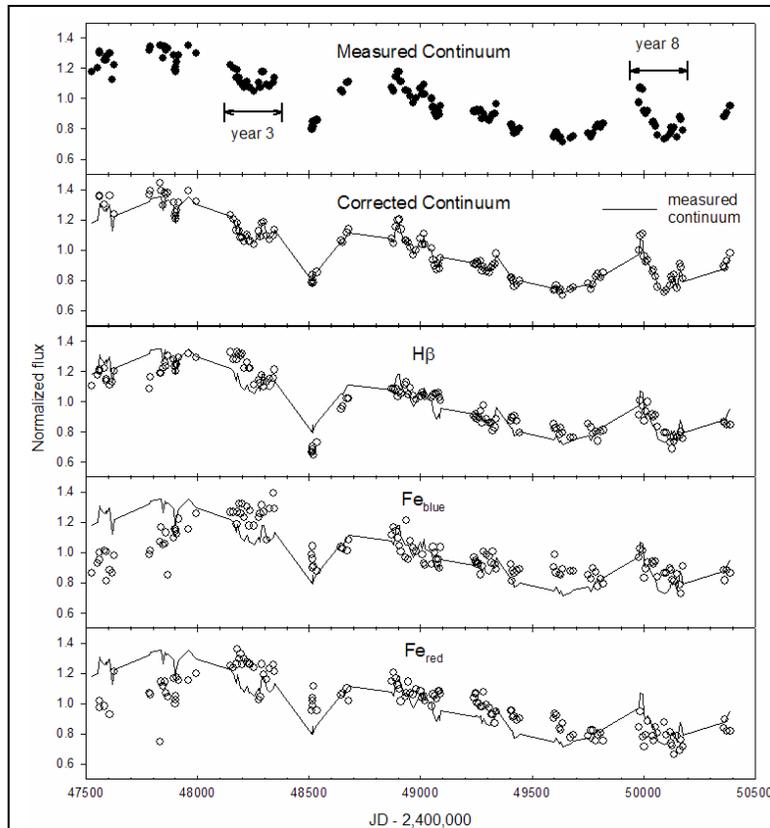

**Figure 3.** Light curves for the 5100Å (c3) continuum point and for the H$\beta$, Fe$_{blue}$ and Fe$_{red}$ emission line fluxes, normalized by their mean values. The solid lines in the lower four panels are the measured continuum light curve, shown for reference. The "Corrected Continuum" points in the second panel down show the (small) effect of correcting the c3 continuum measurements for contamination by the pseudo-continuum due to the Fe II emission bands, using the method explained at the end of §3.

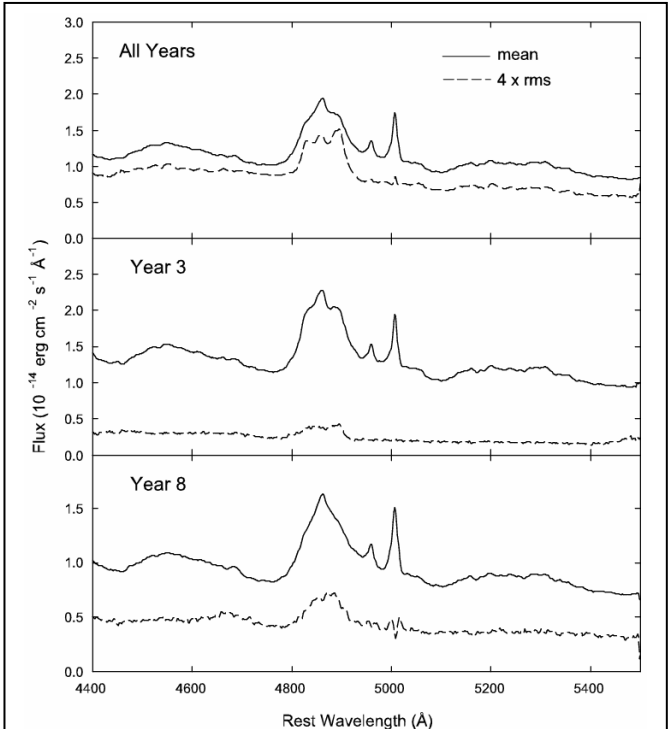

**Figure 4.** Mean and rms spectra for the full data set, and separately for years 3 and 8. The rms spectra have been scaled in flux by a factor of 4, to make them easier to see.

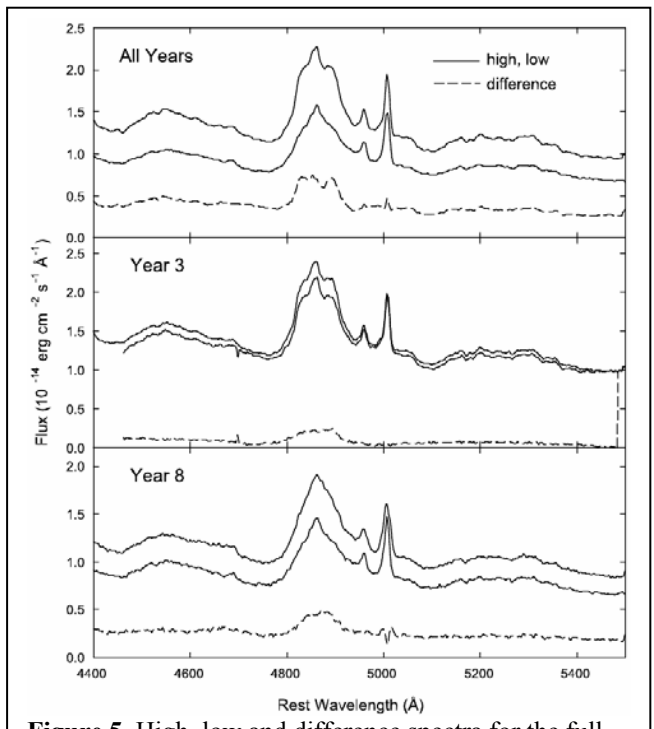

**Figure 5.** High, low and difference spectra for the full data set, and then separately for years 3 and 8.

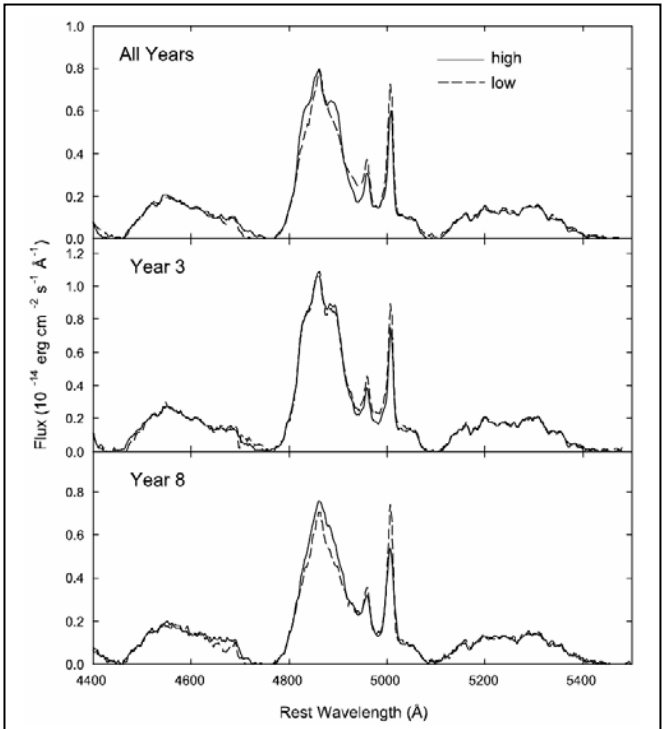

**Figure 6.** High and low-state spectra after the continuum has been subtracted, and then the remainders have been scaled to match at $\lambda 5300$Å. This shows that the shapes of the two Fe II emission bumps have not changed (except in the He II $\lambda 4686$ region for Year 8).

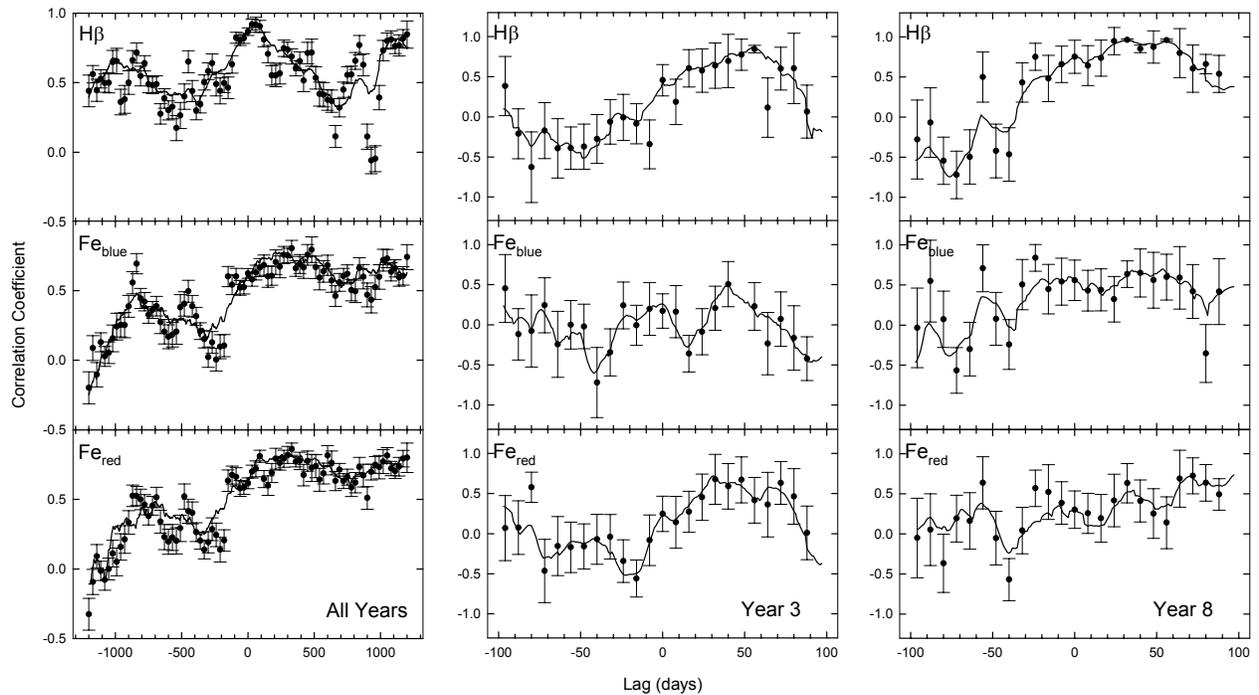

**Figure 7.** Cross-correlation functions for Hβ, Fe$_{blue}$, and Fe$_{red}$ for all years combined, and separately for years 3 and 8. In each frame the interpolation cross-correlation function is shown as a solid line, and the discrete cross-correlation function as individual points with error bars.

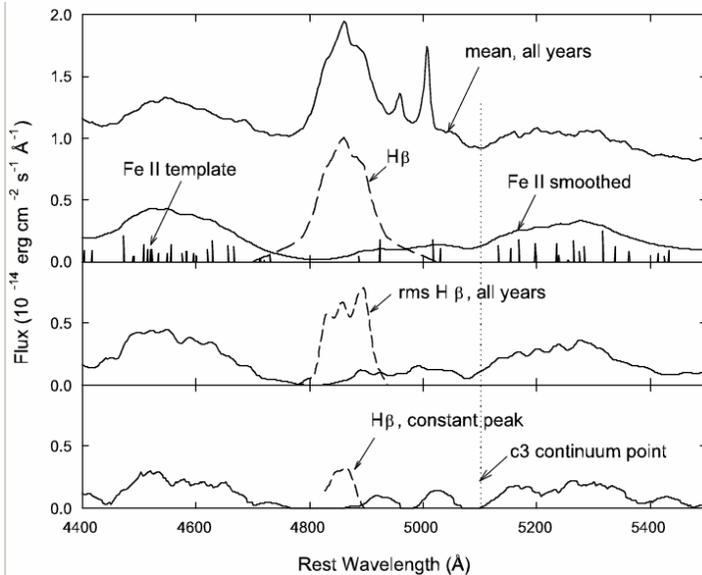

**Figure 8.** The observed mean spectrum for all years (upper solid line in upper panel), compared to examples of smoothed template spectra (remaining solid lines in the three panels). The smoothed template spectra were created by convolving various possible Hβ profiles (dashed lines), created as described in §3, with the Fe II template profile from Véron-Cety et al (2004) which is shown as the series narrow spikes in the upper panel. The "Fe II smoothed" spectrum from the upper panel was used as our estimate of the maximum Fe II contribution to the measured c3 continuum point. The "Corrected Continuum" light curve shown in Figure 3 was computed by scaling this Fe II "pseudo-continuum" contribution in direct proportion to the measured Fe$_{red}$ flux, and then subtracting it from the measured continuum value.

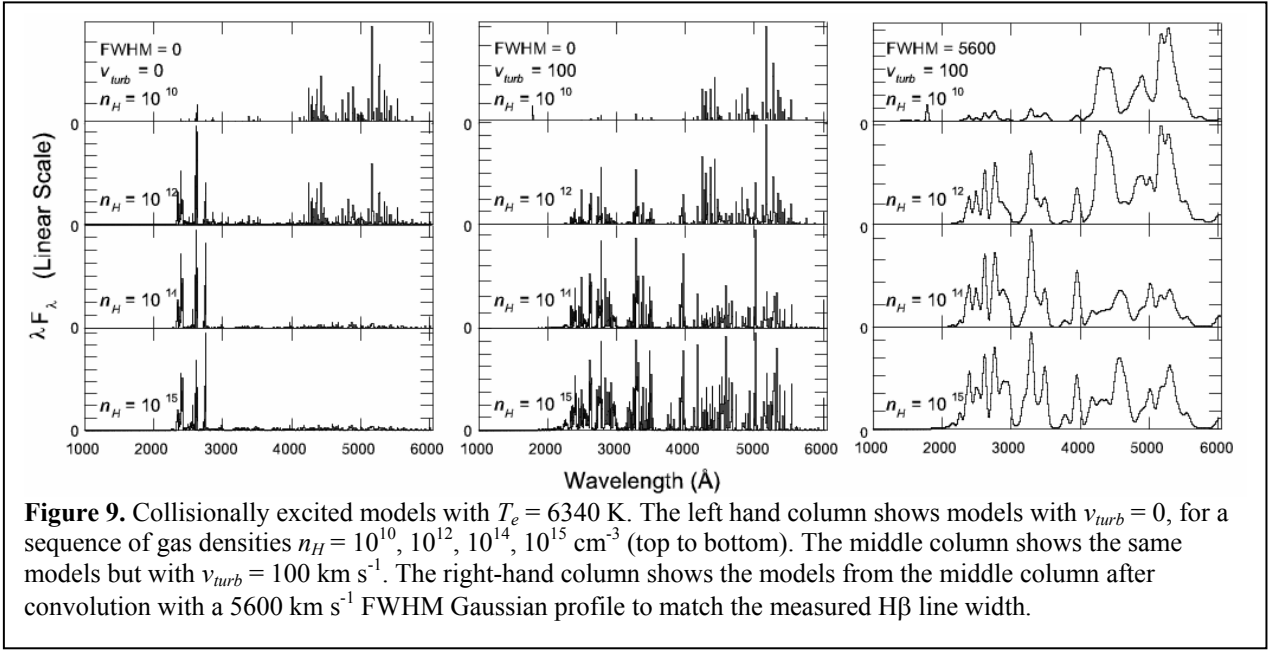

**Figure 9.** Collisionally excited models with $T_e = 6340$ K. The left hand column shows models with $v_{turb} = 0$, for a sequence of gas densities $n_H = 10^{10}$, $10^{12}$, $10^{14}$, $10^{15}$ cm$^{-3}$ (top to bottom). The middle column shows the same models but with $v_{turb} = 100$ km s$^{-1}$. The right-hand column shows the models from the middle column after convolution with a 5600 km s$^{-1}$ FWHM Gaussian profile to match the measured H$\beta$ line width.

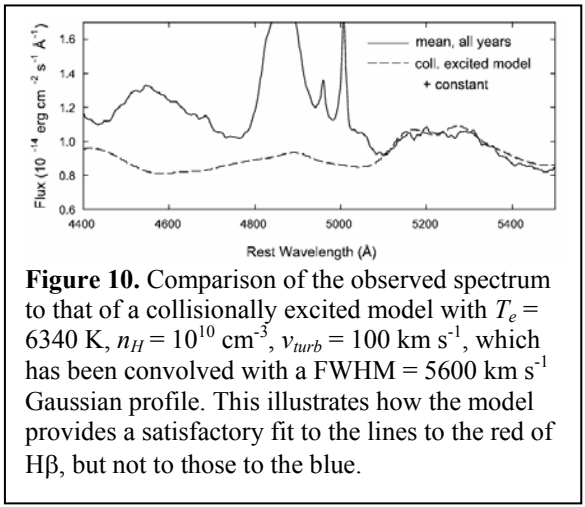

**Figure 10.** Comparison of the observed spectrum to that of a collisionally excited model with $T_e = 6340$ K, $n_H = 10^{10}$ cm$^{-3}$, $v_{turb} = 100$ km s$^{-1}$, which has been convolved with a FWHM = 5600 km s$^{-1}$ Gaussian profile. This illustrates how the model provides a satisfactory fit to the lines to the red of H$\beta$, but not to those to the blue.